\documentclass[twocolumn,showpacs,prb,preprintnumbers,amsmath,amssymb]{revtex4}

\usepackage{dcolumn}
\usepackage{bm}
\usepackage{graphicx}
\usepackage{epsfig}
\usepackage{epsfig}
\usepackage{amssymb}
\usepackage{longtable}

\begin{document}

\title{Hyper-Raman scattering analysis of the vibrations in vitreous boron oxide}

\author{G. Simon, B. Hehlen, R. Vacher, and E. Courtens}

\affiliation{ Laboratoire des Collo\"{\i}des, Verres et Nanomat\'eriaux (LCVN), UMR 5587 CNRS, University of 
Montpellier II, F-34095 Montpellier, France}

\date{\today}

\begin{abstract}

Hyper-Raman scattering has been measured on vitreous boron oxide, $v-$B$_2$O$_3$. 
This spectroscopy, complemented with Raman scattering and infrared absorption, 
reveals the full set of vibrations that can be observed with light. 
A mode analysis is performed based on the local D$_{3h}$ symmetry of BO$_3$ triangles and B$_3$O$_3$ boroxol rings.
The results show that in $v-$B$_2$O$_3$ the main spectral 
components can be succesfully assigned using this relatively simple model.
In particular, it can be shown that the hyper-Raman boson peak arises from 
external modes that correspond mainly to librational motions of rigid boroxol rings. 

\end{abstract}


\maketitle

\section{Introduction}

Group theory provides a complete classification of the vibrational bands of crystals on the basis of symmetry.
This, together with appropriate lattice dynamical calculations or simulations, generally  allows relating the observed
vibrational frequencies to specific atomic motions \cite{Poulet}.
Owing to structural disorder, such assignments are considerably more difficult in glasses. 
The structure of glasses at distances beyond the second atomic neigbhours is hard to unravel using standard structural tools.
However, the structural analysis generally reveals the existence of elementary structural units (ESUs) that are often similar 
to those in the corresponding crystals.
The random network model \cite{Zac32} posits that these ESUs are connected to each other by looser bonds, with 
a broad distribution of interatomic distances and angles.
Connected ESUs can also form larger structural groups that may not exist in the crystal.
One example is given by ring structures, found for example in vitreous silica $v-$SiO$_2$, whose vibrations are observed
in Raman scattering (RS).\cite{Gal82}
It is remarkable that vibrational spectroscopies of glasses mostly reveal well defined modes, pointing to
specific vibrations of ESUs or groups of them.
Structural information can be complemented by the analysis of these vibrations, and there have been
frequent undertakings in that direction.
Thus, the random network description provides a starting point for the analysis of vibrations.

In assigning the vibrations of glasses, a difficulty is that selection rules can be relaxed by either the distortion of the ESUs or by their environment.
This is particularly severe for mixed covalent-ionic glasses.
However, in purely covalent systems it
can already be unsafe to assign modes based on RS or infrared (IR) absorption spectroscopies alone.
For example, it often happens that modes observed in glasses with RS should in fact be inactive on the basis of the simple ESU symmetry.
For this reason, it can be very useful to have results from an additional optical spectroscopy, namely hyper-Raman scattering (HRS).
For example, the rigid librations of SiO$_4$ tetrahedra relating to the low frequency boson peak\cite{Buc86}  
in $v$-SiO$_2$ should be inactive in RS but they have been well identified in HRS\cite{Heh00,Heh02}.
HRS also provides a useful complement to the more usual IR absorption spectroscopy.
IR reflections can be sensitive to the state of the surface \cite{Dum04}, which is hard to control, while HRS probes the bulk. 
Also, the details of the spectral shapes of polar modes derived from IR reflectivity measurements 
can be blurred by the necessary Kramers-Kronig transformation.
An example of that is shown below. 
It turns out that IR active modes can always be directly observed in HRS, with very specific selection rules \cite{Den87}.
As shown in this work, HRS spectroscopy is very helpful for the assignment of the vibrational modes of
vitreous boron oxide, $v-$B$_2$O$_3$.

Boron oxide is an almost ideal glass former which hardly crystallizes at all.
A specificity of $v-$B$_2$O$_3$ is that it contains {\em two} types of ESUs, the triangles BO$_3$ and
the boroxol rings, B$_3$O$_3$, while the single crystal at ambient pressure, which is rhombohedral with $Z$=3,\cite{Gur70}
contains {\em no} boroxol rings.
For this reason, observing vibrations in the crystal would be of very limited guidance to assign the vibrations in the glass. 
The existence of boroxol rings in the glass is known from 
x-ray diffraction \cite{Moz70}, NMR \cite{Jel77}, RS \cite{Gal80}, and neutron scattering \cite{Joh82}.
Although boroxols result from the bonding of three BO$_3$ triangles, these ESUs are special in that their internal B--O--B angle is
$\cong 120 ^\circ$, while it is significantly larger for O atoms not belonging to boroxols as discussed below.
The specific vibrations of boroxols have been the subject of several studies \cite{Tos90}.
However, there has been a very long debate regarding the relative abondance of these rings which
could not be ascertained from structure factors alone.
It is only recently, with the help of NMR spectroscopy \cite{Zwa05} and of simulations of the vibrational spectra \cite{Uma05}  
that it was established that approximately 3 out of 4 boron atoms belong to B$_3$O$_3$ rings in $v-$B$_2$O$_3$, 
although this result was further challenged \cite{Swe06,Uma06}.

\begin{figure*}[t]
\includegraphics[width=14cm]{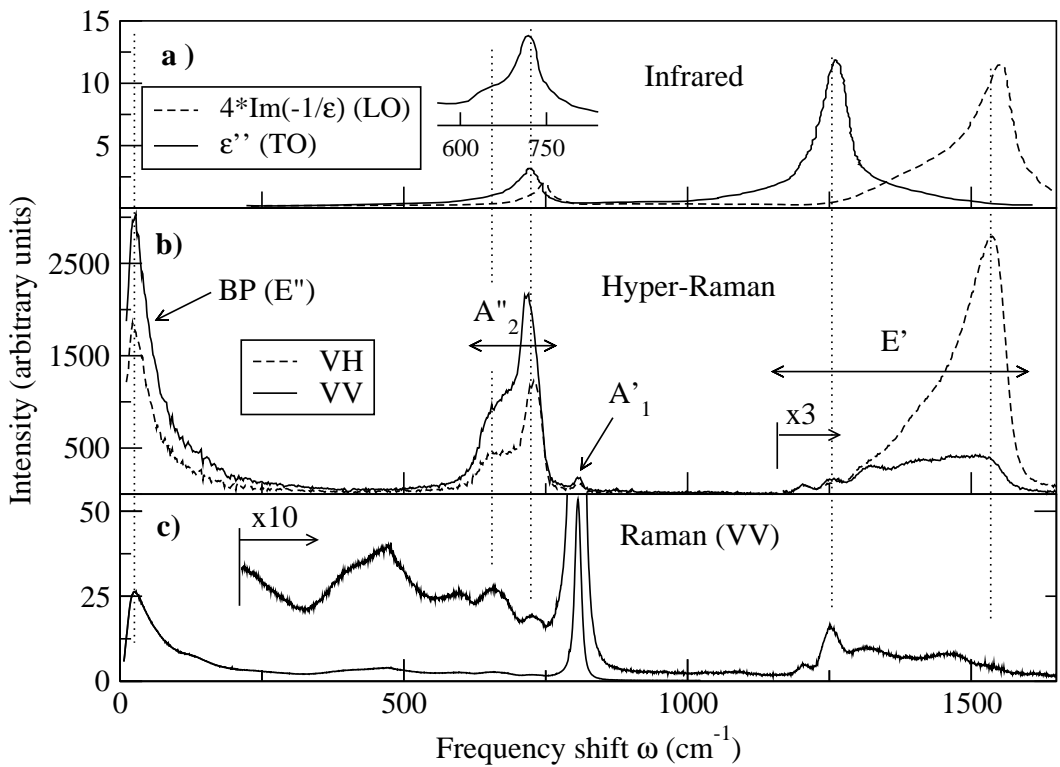}
\caption{\label{spectra} Spectroscopy of $v$-B$_2$O$_3$: {\bf a)} IR absorption from TO 
(-\;-) and LO modes ({\bf---}) derived from IR reflectivity measurements \cite{Gal80}.
The inset shows the lowest frequency polar band measured in transmission \cite{Ten72} which is more structured than the TO peak 
derived from reflection;
{\bf b)} VV and VH hyper-Raman spectra obtained in 90$^\circ$ scattering; 
{\bf c)} Raman (VV) spectrum from the same sample as in b). The frequencies are expressed in wavenumbers, $\omega /2\pi c$.}
\end{figure*}

In this paper we report the complete spectroscopy of $v$-B$_2$O$_3$ performed with both hyper-Raman scattering (HRS) and RS.
HRS is a non-linear optical spectroscopy in which two incident photons of angular frequency $\omega$ combine with one excitation 
of the material at $\omega _ {\rm v}$ to produce one scattered photon at $2\omega \pm \omega _ {\rm v}$.\cite{Den87}
The lower sign corresponds to a Stokes event (creation of a material excitation), while the upper sign is for an anti-Stokes 
event.
RS was performed on the same sample in order to have fully comparable data.
One interest in combining HRS with RS is that the selection rules of these two spectroscopies are often complementary, modes
forbidden in RS being mostly allowed in HRS, and vice versa.
Both longitudinal optic (LO) and transverse optic (TO) modes are active in HRS, with selection rules that allow separating them \cite{Den87}.
The results presented below will be compared to IR data from the literature.
One can then identify clearly the TO and LO contributions.
Spectra of pure $v$-B$_2$O$_3$ obtained in IR, RS, and HRS are presented in Fig. \ref{spectra}. 
The RS and HRS spectra are remarquably dissimilar.
The polarized (VV) and depolarized (VH) HRS spectra exhibit three main bands plus a weak line around 808 cm$^{-1}$. 
The three intense bands are the boson peak (BP) at low frequencies, and TO-LO responses centered around 700 cm$^{-1}$ and 1400 cm$^{-1}$. 
The latter are immediately identified as polar vibrations since they are strongly active in IR.
The weakest HRS line around 808 cm$^{-1}$ is the vibration that dominates the polarized RS spectrum. 
The BP, below $\sim 200$ cm$^{-1}$, is the only strong feature observed both in RS and in HRS.\\ 
\indent
The paper is organized as follows:
in Section II, the experimental aspects are described, including the spectrometer and the characterization of the sample;
the structural model used for the description of the data is described in Section III; 
the analysis of the IR, RS, and HRS spectra is performed in sections IV, V, and VI, for the non-polar, 
polar, and  boson-peak modes, respectively; finally, a discussion concludes the paper in Section VII.

\section{Experimental aspects}

The HRS efficiency is usually very small, about 10$^6$ times weaker than the RS one. 
This explains that this spectroscopy was essentially restricted in the past to favorable nonlinear solids,
such as ferroelectric oxides.
Several recent instrumental  advances, such as high power pulsed lasers, multichannel photo-counting devices, 
and large aperture and aberration corrected optical design, opened up the possibility for many new HRS investigations, 
particularly in glasses. 
In our setup, the hyper-Raman spectrum is excited by a diode pumped Q-switched YAG-laser emitting at the wavelength
$\lambda$ = 1064 nm.\cite{laser} 
For the experiments in B$_2$O$_3$, we used typical repetition rates of 2500 Hz, leading to $\sim$ 20 ns pulse width. 
As this glass is very robust and optically transparent at this wavelength, it was possible to perform measurements with incoming 
peak powers up to $\sim$ 35 kW without alteration of the sample. 
The beam is focused with a $f$=5 cm lens to a $\sim$ 20 $\mu$m diameter waist.
The scattered light is collected by a $f/1.5$ photo-objective either in near forward-, in 90$^\circ$-, or in back-scattering geometry. 
The polarization of the incident light is vertical (V), {\em i.e.} perpendicular to the scattering plane.
That of the scattered light can be selected either V or H (horizontal) with an analyser that combines a half-wave plate with
a wide band and large aperture Glan polarizer. 
The latter is fixed at V and the choice of the polarization V or H is done by 45$^\circ$-rotation of the $\lambda/2$-plate.
Particular attention was given to accurate polarization measurements. 
To this effect, the setup was calibrated with the strongly polarized Raman line of CCl$_4$ at $\approx 460$ cm$^{-1}$.\cite{Mur67}
This allowed estimating that the polarization leakage of our setup is below  0.1 \%, at least up to that frequency range. 
The HRS spectra are dispersed with a Jobin-Yvon FHR-640 single-grating diffractometer of $f/5.6$ aperture, 
and detected with a nitrogen cooled CCD camera \cite{CCD}.  
The dispersion of 2.4 nm/mm of the 600 grooves/mm grating allows covering a spectral range of $\simeq$ 2500 cm$^{-1}$ 
on the detector  
This generally contains the full HRS spectrum.
It is then acquired with a resolution of $\sim$ 6 cm$^{-1}$ (FWHM) using an entrance slit of 100 $\mu$m. 
A second grating with 1800 grooves/mm is also available for high resolution experiments.
It zooms on a spectral range of about 800 cm$^{-1}$, with a spectral resolution of $\sim$ 2 cm$^{-1}$ (FWHM) at a 100 $\mu$m entrance slit. 
RS spectra of the same B$_2$O$_3$ sample were obtained with 514.5 nm excitation and analysed with a Jobin-Yvon T64000 
triple-grating monochromator.
We remark in passing that an attractive property of HRS in glasses, as compared to RS, is the inelastic over elastic signal ratio
which is orders of magnitude more favorable in HRS, about 10$^8$ times larger.
This provides a much better contrast at low frequencies, in particular in the BP region.
In the high resolution single-grating mode, the tail of the hyper-Rayleigh line is generally negligible above 10 cm$^{-1}$,
while RS would be very poluted by the elastic line in such a configuration.  

The $v$-B$_2$O$_3$ sample was prepared from isotopically pure (99.6\%) $^{11}$B$_2$O$_3$ containing $\sim$ 1 wt\% moisture. 
The material was heated to 1100 $^\circ$C in a platinum crucible, quenched on a heat
conducting plate, and annealed at 571 $^\circ$C 
for 200 hours. Its water content, $\sim$ 0.8 wt\%, was measured by infrared transmission of a thin slice.\cite{Ram97} 
When required, $v$-B$_2$O$_3$ data were complemented by HRS from two lithium borate glasses, 
4B$_2$O$_3$:Li$_2$O and 2B$_2$O$_3$:Li$_2$O, kindly provided by Dr. A. Matic 
from Chalmers University of Technology, G\"oteborg, Sweden.

\section{The structural model} 

The structure of vitreous boron oxide consists of a network of quasi-planar BO$_3$ triangles connected by their apical oxygens. 
Three triangles can associate to form a planar boroxol ring B$_3$O$_3$. 
These two elementary structural units (ESUs), lone triangles and boroxol rings, are well defined,
as seen from neutron diffraction \cite{Joh82} and simulations \cite{Uma05} results.
The angular distribution of the O-B-O bonds is 120$^\circ$ $\pm$ 3.3$^\circ$ in triangles,\cite{Uma05} showing that they are rather flat. 
The boroxols are even closer to the perfect symmetry, as shown in Ref. \cite{Uma05}. 
The angular distribution of the B-O-B bonds bridging two ESUs is $\sim$130$^\circ$ from 
Ref. \cite{Moz70,Han94} or $134.4 ^\circ \pm  9.2^\circ$ from Ref. \cite{Uma05}.
These have a much larger spread and they are also not restrained to the ESU plane, leading to the three dimentional
random structure of the glass.

The spectra will be indexed according to a simple model in which only two types of ESUs
are considered, the B$_3$O$_3$ boroxol {\em rings} and the BO$_3$ {\em triangles} (Fig. 2).
Since the sum of one ring plus one triangle gives exactly two formula units, 2(B$_2$O$_3$),
our model should contain the required number of modes if the glass would be composed exclusively of triangles and rings in the 
ratio 1:1.
Alternatively, one could define the units as BO$_{3/2}$ and B$_3$O$_3$O$_{3/2}$, in the spirit of the split atom model\cite{Gid93}.
This has the merit to show that the ESUs transform as the equilateral triangles formed by the three O$_{1/2}$ splits atoms. 
However, this would  be inconvenient for realistic calculations of the internal ESU vibrations, and it will not be adopted here.
Our structural description is certainly very simplified.
Strictly speaking, it only allows each triangle or ring to connect with three rings or triangles, respectively, while the real glass could  
contain interpenetrating  ring networks, $m($B$_3$O$_3$ + $-{\rm O}-$), and triangle networks, B$_n$O$_{2n+1}$. 
If less than 75\% of boron atoms belong to rings, the glass will {\it necessarily} form other ESUs that preserve the 
stoichiometry, such as B$_2$O$_5$, B$_3$0$_7$, etc. 
Considering the dynamical properties, our model should nicely reproduce the vibrations of rings.
However, it completely omits the $-{\rm O}-$ bonds between two connected rings, and it can only 
partly index the B$_n$O$_{2n+1}$ motions in terms of vibration of BO$_3$ units.
\begin{figure}
\includegraphics[width=5cm]{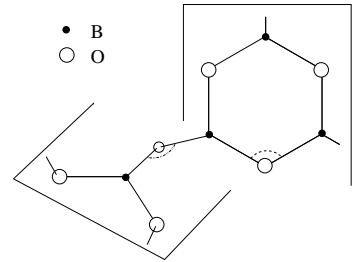}
\caption{Schematic assembly of a ring and a triangle in $v$-B$_2$O$_3$.
These two ESUs form two B$_2$O$_3$ molecules. The relative orientation of the planes 
containing the triangle and the ring is random. It is the repetition 
of such patterns that produces the disordered network of $v$-B$_2$O$_3$ in this model.
The possibility that occasionally two boroxols be directly connected with one --O-- atom, as in Fig. 11c below, is not excluded. However in
that case there should also be two directly connected triangles, --B$_2$O$_5$--, for compensation. }
\end{figure}
 
\begingroup
\squeezetable
\begin{table*}[t]
\caption{Selection rules of modes in the point group D$_{3h}$, with the matrix elements for 
the Raman tensors\cite{Poulet}, $\alpha^\zeta_{ij}$, and the symmetric hyperpolarisability tensors\cite{Den87}, $\beta^\zeta_{ijk}$.
The latter are shown in the contracted indices notation.\cite{Indices}
The last column displays the eigenvectors of the normal modes for the flat CO$_3$ radical of calcite \cite{Poulet}.
T and R designate the 3 external translations and rotations respectively, the so-called
non-genuine normal vibrations for isolated molecules. \cite{Her45}}  
\begin{tabular}{cccc|c}\hline \hline
  & & & &  \\
 Mode & \parbox{3cm}{IR} & \quad \parbox{4cm}{RS\cite{Poulet}} & \parbox{3cm}{HRS\cite{Den87}} &  
 \parbox{5cm}{Normal modes\cite{Poulet,Her45}} \\ 
 & & & &  \\ \hline
 & & & &  \\
 A$'_1$ &        &  
 $\left | \begin{tabular}{ccc}
 a & $\cdot$ & $\cdot$ \\ 
$\cdot$ & a & $\cdot$ \\
$\cdot$ & $\cdot$ & b \\
\end{tabular} \right | $ & 
$\left \| \begin{tabular}{cccccc}
c & -c & $\cdot$ & $\cdot$ & $\cdot$  & $\cdot$ \\
$\cdot$ & $\cdot$ & $\cdot$  & $\cdot$  & $\cdot$ & -c \\
$\cdot$ & $\cdot$ & $\cdot$  & $\cdot$  & $\cdot$ & $\cdot$
\end{tabular} \right \| $ & 
\parbox{1.5cm}{{\epsfysize=1.5cm\epsffile{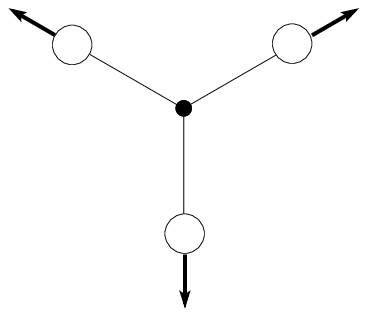}}}\\ \hline
& & & &  \\ 
E' & ($\mu_x,\ \mu_y$) & 
\parbox{2cm}{$\left | \begin{tabular}{ccc}
 d & $\cdot$ & $\cdot$ \\ 
$\cdot$ & -d & $\cdot$ \\
$\cdot$ & $\cdot$ & $\cdot$  \\
\end{tabular} \right | $\\ 
\ \\
\ \\
 $\left | \begin{tabular}{ccc}
 $\cdot$ & -d & $\cdot$ \\ 
-d & $\cdot$ & $\cdot$ \\
$\cdot$ & $\cdot$ & $\cdot$  \\
\end{tabular} \right | $\\ \ \\ \ \\} & 
\parbox{2cm}{
$\left \| \begin{tabular}{cccccc}
3e & e & f & $\cdot$ & $\cdot$  &  $\cdot$  \\
$\cdot$ & $\cdot$  & $\cdot$ & $\cdot$  & $\cdot$  & e  \\
$\cdot$ & $\cdot$ & $\cdot$  & $\cdot$  & f & $\cdot$
\end{tabular} \right \| $ \\
\ \\
\ \\
$\left \| \begin{tabular}{cccccc}
$\cdot$ & $\cdot$ & $\cdot$ & $\cdot$ & $\cdot$  &  e  \\
e & 3e & f & $\cdot$  & $\cdot$  & $\cdot$ \\
$\cdot$ & $\cdot$ & $\cdot$  & f  & $\cdot$ & $\cdot$ 
\end{tabular} \right \| $ \\ \ \\ \ \\} & 
\parbox{6cm}{
\parbox{3.5cm}{\epsfysize=2.6cm\epsffile{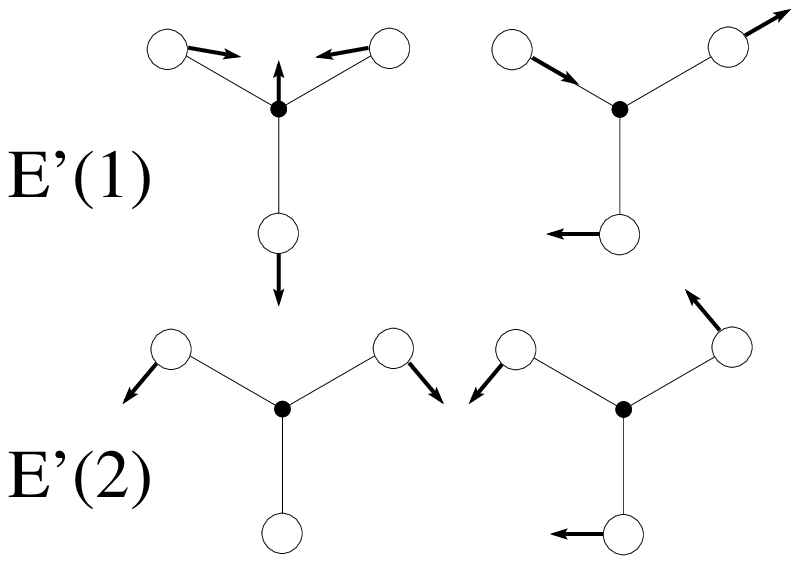}}\ \ , (T$_x$, T$_y$)} \\ \hline
& & & &  \\ 
A$^"_2$ &  $\mu_z$ &   &
 $\left \| \begin{tabular}{cccccc}
$\cdot$ & $\cdot$ & $\cdot$ & $\cdot$ & b  &  $\cdot$ \\
$\cdot$ & $\cdot$ & $\cdot$ & b  & $\cdot$  & b \\
b & b & a  & $\cdot$  & $\cdot$ & $\cdot$
\end{tabular} \right \| $ & 
\parbox{2.9cm}{{\epsfxsize=1.8cm\epsffile{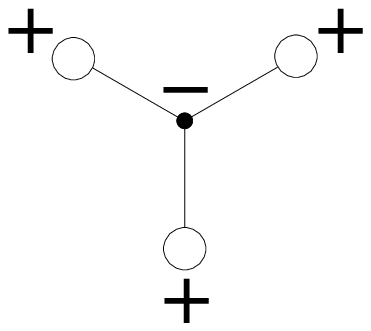}}}\ \ , T$_z$ \\ \hline
& & & &  \\ 
A$'_2$  &  &  & 
$\left \| \begin{tabular}{cccccc}
$\cdot$ & $\cdot$ & $\cdot$ & $\cdot$ & $\cdot$  &  -d \\
-d  & d & $\cdot$ & $\cdot$  & $\cdot$  & $\cdot$ \\
$\cdot$  & $\cdot$ & $\cdot$ & $\cdot$  & $\cdot$ & $\cdot$ 
\end{tabular} \right \| $ & R$_z$\\ 
& & & &  \\ \hline
 & & & &  \\
E''   &    & 
\parbox{2cm}{
$\left | \begin{tabular}{ccc}
 $\cdot$ & $\cdot$ & $\cdot$ \\ 
$\cdot$ &  $\cdot$ & c \\
$\cdot$ & c & $\cdot$  \\
\end{tabular} \right | $\\
\ \\ 
\ \\
$\left | \begin{tabular}{ccc}
 $\cdot$ &  $\cdot$ & -c \\ 
 $\cdot$ & $\cdot$ & $\cdot$ \\
-c & $\cdot$ & $\cdot$  \\
\end{tabular} \right | $ }& 
\parbox{3cm}{
$\left \| \begin{tabular}{cccccc}
$\cdot$ & $\cdot$ & $\cdot$ & b & $\cdot$  & $\cdot$\\
 $\cdot$ & $\cdot$ & $\cdot$ & $\cdot$ & b  & $\cdot$ \\
$\cdot$  & $\cdot$ & $\cdot$ & $\cdot$  & $\cdot$ & b
\end{tabular} \right \| $ \\
\ \\
\ \\
$\left \| \begin{tabular}{cccccc}
$\cdot$ & $\cdot$ & $\cdot$ & $\cdot$ & -b  & $\cdot$\\
 $\cdot$ & $\cdot$ & $\cdot$ & b  & $\cdot$  & $\cdot$ \\
-b  & b & $\cdot$ & $\cdot$  & $\cdot$ & $\cdot$
\end{tabular} \right \| $ }& 
\parbox{1cm}{R$_x$ 
\\ \ \\ \ \\ \ \\ R$_y$}\\
& & & & \\ \hline\hline
\end{tabular}
\end{table*}
\endgroup
Regarding the symmetry properties, both BO$_3$ triangles and B$_3$O$_3$ boroxol rings belong to the point group D$_{3h}$. 
Hence, the vibrations decompose into the following irreducible representations :
\begin{eqnarray}
{\rm [1,2]\; A'_1 + [1,2]\; A'_2 + [2,2]\; A''_2 + 
		\rm [3,4]\; E' + [1,2]\; E'' }\nonumber\\
		\hfill 		 
\end{eqnarray}
The figures within square brackets show the number of independent 
representations for triangles and rings, respectively.
The IR, RS and HRS activities of each type of modes are sumarized in Table I. 
These selection rules, derived for perfect symmetry,\cite{Cyv65} do not 
account for possible distortions of the ESUs. 
In spite of  the above mentioned limitations, we show in the following that the selection
rules in Table I match remarkably well the experimental observations, 
thus allowing for a quantitative description of most vibrations of $v$-B$_2$O$_3$. 

It is useful to provide a brief phenomenological description of the interaction of the ESUs with light.
Light-scattering spectra are given by the Fourier transform of the space and time correlation functions 
of the dipole-moment fluctuations $\delta {\bf p}$ associated with vibrations and induced in the material by an incident 
electric field {\bf E} $ \propto e^{-i \omega _i t}$. 
For a structural unit $m$, the {\it induced} dipole ${\bf p}^m$ can be expressed in terms of the local field ${\bf E}^l$ as
\begin{equation}
{\bf p}^m =  {\bm \chi}^{(1)}_m \cdot {\bf E}^l + \frac{1}{2}{\bm \chi}^{(2)}_m : {\bf E}^l {\bf E}^l + ...\quad , 
\end{equation}
where  ${\bm \chi}^{(1)}_m$ and ${\bm \chi}^{(2)}_m$ are the first and second-order 
polarizability tensors, respectively.
Proper consideration of the local field is important for quantitative evaluations \cite{Mun01,Mun02}.
However, for what symmetries are concerned, we assume that ${\bf E}^l$ is essentially proportional to ${\bf E}$ in a glass.
It is the modulation of the polarizabilities by the normal modes of amplitude $W^\zeta \propto e^{\mp i \omega_\zeta t}$ for mode 
$\zeta$, which leads to a modulation of ${\bf p}^m$ at the Raman and hyper-Raman frequencies.
One thus defines the molecular Raman and hyper-Raman polarizability-derivative tensors, 
${\bm \alpha}^{\zeta,m} = (\partial{\bm \chi}^{(1)}_m/\partial W^\zeta)$ and
${\bm \beta}^{\zeta,m} = (\partial{\bm \chi}^{(2)}_m/\partial W^\zeta)$, 
respectively.
In addition, if the dipole moment ${\bm  M}^m$ of the ESU $m$ is modulated by the mode $\zeta$, there is infrared absorption at
$\omega _\zeta $, with an amplitude proportional to ${\bm \mu}^{\zeta,m} = (\partial{\bm  M}^m/\partial W^\zeta)$.
The properties of the tensors ${\bm \mu}^{\zeta,m}$, ${\bm \alpha }^{\zeta,m}$, and ${\bm \beta }^{\zeta,m} $, have been 
tabulated for all  molecular symmetries, {\em e.g.} in Ref. \cite{Cyv65}.
Table I reproduces the results for the D$_{3h}$ point group. 
In the Table, the orthogonal frame $(x,y,z)$ is fixed to the ESU, with 
$z$ perpendicular to the ESU plane, and $x$ joining the center of the triangle to one of its vertices. 
The tensors in that frame will be written ${\bm \alpha }^{\zeta}$ and ${\bm \beta }^{\zeta}$.
The vibrational eigenvectors for a AX$_3$ planar molecule are sketched in the last column of Table\,I.

As the two incident photons are at the same laser frequency $\omega _i$, 
the hyper-Raman polarization fluctuation, $\delta {\bf p}^{\zeta,m}={\bm \beta}^{\zeta,m} : {\bf E}^l {\bf E}^l \; W^\zeta $  
leads to a scattered field at frequency $\omega_s = 2\omega_i \pm \omega_ \zeta$.
In this case the tensor $\bm \beta $ is obviously symmetric with respect to its last two indices, $\beta _{ijk} = \beta _{ikj}$.
Moreover, since both $\omega _i$ and $\omega _s$ are far from any material 
resonance, ${\bm \beta} ^\zeta $ can be approximated as fully symmetric in all permutations of its three cartesian indices \cite{Den87}. 
The corresponding scattering vector {\bf q} is given by
${\bf q} = \pm ({\bf k}_s - 2{\bf k}_i)$, where ${\bf k}_s$ and ${\bf k}_i$ are the wave vectors of the scattered and incident
radiation, respectively. 

The integrated HRS intensity for mode $\zeta$, projected onto a unit polarization vector $\bf e$,
is proportional to the square of the corresponding polarization density.
The latter is given by a sum over the $m$ ESUs vibrating at $W^\zeta $ within a unit volume, 
\begin{equation}
I^\zeta_{\rm HR} \propto  |\sum_{m} {\bf e } \cdot \delta{\bf p}^{\zeta,m} | ^2 \quad .
\end{equation}
In the case of B$_2$O$_3$, our model implies that the sum extends only over triangles, or only over rings, depending on the particular mode $\zeta $.

\section{Hyper-Raman scattering from non-polar modes}

When each ESU vibrates independently from the others, the cross terms in Eq. 3 vanish on the average, 
and the summation can be extracted from the square. 
Similarly to the Raman case, one obtains then for the {\it incoherent} 
hyper-Raman scattering intensity,
\begin{equation}
I^{\zeta,\;{\rm inc}}_{\rm HRS} \propto \sum_{m} \left |
{\bf e} \cdot {\bm \beta}^{\zeta,m} : {\bf E\,E} \right |^2 \quad .
\end{equation}
Eq. (4) generally applies to non-polar {\it internal} vibrations.
The scattering is then ``local'' and thus independent of {\bf q}.
Implicit in Eq. (4) is the fact that the polarization vectors are fixed in the laboratory frame, while in a glass different ESUs have
different orientations in that frame.
Hence, Eq. (4) implies an angular averaging.
The explicit calculation of (4) involves three rotation matrices in each term, to bring ${\bm \beta }^{\zeta }$ from the table to the
particular ${\bm \beta }^{\zeta ,m}$ for each ESU $m$,
or conversely and more conveniently, to rotate {\bf E} and ${\bf e}$ to the ESU frame.
For a macroscopically isotropic medium this averaging is uniform, leading to results
which are found {\em e.g.} in Ref. \cite{Ber66}.
One defines a depolarization ratio for mode $\zeta$,
\begin{equation} 
\rho^{\zeta,\;{\rm inc}}_{\rm HRS} = {\rm I}^{\zeta,\;{\rm inc}}_{\rm VH} \; / \; {\rm I}^{\zeta,\;{\rm inc}}_{\rm VV} \quad ,
\end{equation}
where VV stands for ${\bf e} \parallel {\bf E}$, while VH means ${\bf e} \perp {\bf E}$.
For the local D$_{3h}$ symmetry of B$_2$O$_3$, one has 
$\rho^{\zeta,\;{\rm inc}}_{\rm HRS} = 2/3$ for all non-polar modes \cite{Ber66}, a result 
which will be important in the mode analysis below, as it allows distinguishing polar excitations from incoherent contributions. 

There is at least one feature in the HRS spectra that can be unambiguously associated with a non-polar vibration. 
It is the relatively weak line at 808 cm$^{-1}$ on Fig. 1{\bf b} which gives 
the narrow peak dominating the Raman spectrum (Fig. 1{\bf c}).  
This mode was assigned long ago to radial breathing motions of boroxol rings \cite{Win82}.
The proposal that it might relate to a coherent symmetric stretch of the bridging oxygens over the entire network \cite{Mar81} should 
be abandoned in this particular case, since the large spread in -B-O-B- angles outside boroxol rings could not produce such a narrow width.
Raman scattering from isotopically pure samples containing $^{10}\rm B$, $^{11}\rm B$, $^{16}\rm O$, and $^{18}\rm O$, has established beyond
reasonable doubt that this mode is associated with a local symmetric stretch A$'_1$ within boroxols \cite{Win82}. 
It is the remarkable regularity of the boroxols that accounts for the narrowness of this vibration.
The eigenvectors, sketched in the inset of Fig. 3, are completely dominated by the oxygen motion, as confirmed
by isotopic substitution \cite{Win82}.
Incidentally, this explains that this motion essentially does not couple different boroxol rings.
The vibration is non-polar, and thus inactive in IR, as observed in Fig. 1.   
From Table I, it is also expected to give intensity both in RS and HRS which is indeed confirmed by the experiment.
In fact, it is the same transition matrix elements that are involved in the microscopic calculation of 
$\alpha _{xx}$ and $\beta _{xxx}$.\cite{Den87}
Therefore there is no reason to believe that the HRS activity should be anomalously small compared to the RS activity.
Hence, the relative size of the A$'_1$ peaks observed in the two spectroscopies does provide a scale for the comparison of the spectral intensities.
The relatively large strength of the other features observed in HRS must derive from the coherent enhancement of these 
signals\cite{Sim06}, as explained below.
\begin{figure}[h]
\includegraphics[width=8cm]{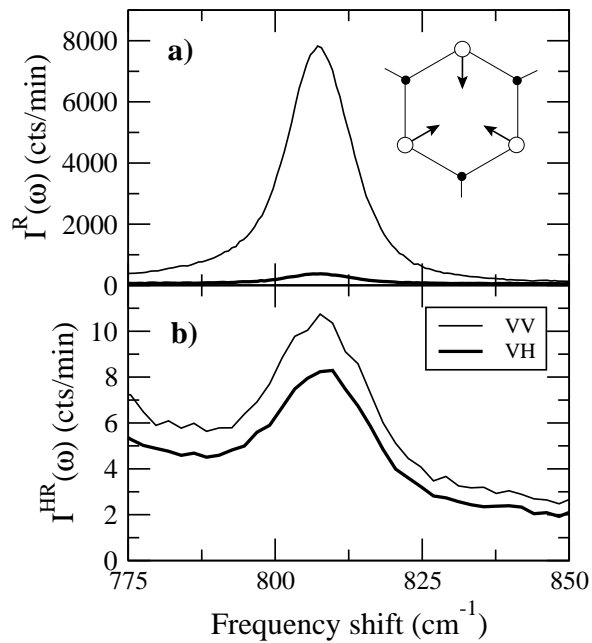}
\caption{Polarization  measurement of 
the breathing mode of boroxol rings in $v$-B$_2$O$_3$ 
(A$'_1$ symmetry): {\bf a)} RS, and {\bf b)} HRS.} 
\end{figure} 

Figure 3 shows the detail of our Raman and hyper-Raman  polarized (VV) and depolarized (VH) spectra.
It is already known that the RS signal is strongly polarized for this particular mode.
A depolarization  ratio $\rho^{A'_1} _{\rm RS} = {\rm I}_{\rm VH} / {\rm I}_{\rm VV} = 1/28$ was reported in Ref. \cite{Gal80}.
In our most reliable measurement shown in Fig. 3a, we find $\rho^{A'_1} _{\rm RS} = 1/23$.
All normal modes of D$_{3h}$ symmetry have a RS depolarization ratio equal to 3/4, except for A$'_1$ modes.
The depolarization ratio of  A$'_1$ is different as there is an additional contribution to ${\rm I}_{\rm VH}$
arising from $\alpha _{xx} \neq \alpha _{zz}$.
The inverse ratio is then given by
$$ \frac{I_{\rm VV}}{I_{\rm VH}}=\frac{1}{\rho_{\rm RS}^{A'_1}}=\frac{5}{3} 
\left ( \frac{2\alpha_{xx}+\alpha_{zz}}{\alpha_{xx}-\alpha_{zz}}\right )^2
+\frac{4}{3}\ ,$$ 
leading to strongly polarized lines as soon as $\alpha_{xx}\simeq\alpha_{zz}$.
For a depolarization ratio equal to 1/23 there are two possiblities : either $\alpha_{xx}=2.86\,\alpha_{zz}$ or 
$\alpha_{xx}=0.46\,\alpha_{zz}$.
With the covalent bonds in plane, the former seems more likely.
In HRS we find $\rho^{A'_1}_{HRS}\simeq$ 0.67$\pm$ 0.015 (Fig. 3b), 
in very good agreement with the theoretical value of 2/3. 
Interestingly, we find that this mode perfectly fulfills the selection rules for incoherent scattering.
This certainly arises  from the very regular structure of the boroxol rings 
in the glass and from the fact that only the oxygen atoms move. 
It is now of interest to investigate whether the selection rules, which apply so well for $\zeta=$A$'_1$, are also
satisfied for the other boroxol vibrations.

\section{Hyper-Raman scattering from polar modes}

In situations where several units vibrate with a fixed phase relationship, there is {\it coherence}, and the cross products in the
development of Eq. 3 cannot be neglected.
In HRS, this occurs in two important cases :
($i$) for {\it polar vibrations}, where the dipole-induced electric field imposes an extended phase relationship in the material, and
($ii$) for {\it external modes}, {\it e.g.} for librations in which adjacent units move together like cogwheels, also leading to a certain degree of coherence.
The intensity in Eq. 3 can be separated into two terms,
\begin{equation}
I_{HR}^{\zeta} \propto 
 \sum_{m} | \delta{\bf p}^{\zeta,m} | ^2 + 
\sum_{m\neq n} \delta{\bf p}^{\zeta,m} \cdot \delta{\bf p}^{\zeta,n} \quad .
\end{equation}

\subsection{Selection rules} 

The important point is that the scattered intensity for modes 
of type ($i$) or ($ii$) is affected by the coherent term on the right hand side of Eq. 6. 
For isotropic ({\em i.e.} incoherent) averaging, the sum over 
$ \delta{\bf p}^{\zeta,m}\cdot \delta{\bf p}^{\zeta,n}$ in Eq. 6 is zero.
It is the existence of correlations between the units $m$ and $n$ that gives weight to this term.
For polar modes, the correlation function depends on the symmetry of the vibrations and on the 
the polarization field. Its calculation is beyond the scope of this paper.
A device to account for this anisotropic averaging is to separate ${\bm \beta}^\zeta$ in two terms, 
${\bm\beta}^\zeta = \Delta{\bm\beta}^\zeta + \bar{\bm\beta}^\zeta$.\cite{Den83} 
$\Delta{\bm\beta}^\zeta$ corresponds to the local scattering process and has the symmetry specified in 
Section III, while $\bar{\bm\beta}^\zeta$ corresponds to a modulation over the average $\infty\infty$m glass symmetry.
The properties of the latter are condensed into a single matrix\cite{Den83} : 
\begin{eqnarray}
\bar{\bm\beta}^\zeta = 
\begin{tabular}{cc}
\hspace{-.6cm} XX \  \ YY \ \ \ ZZ \ & \\ 
\begin{tabular}{|ccc|c} 
\ $3a^\zeta$ \ &\  $a^\zeta$ \ &\  $a^\zeta$ \ &\ X\\  
\ $a^\zeta$ \ &\  $3a^\zeta$ \ &\  $a^\zeta$ \ &\ Y\\ 
\ $a^\zeta$ \ &\  $a^\zeta$ \ &\  $3a^\zeta$ \ &\ Z  
\end{tabular} 
\end{tabular}
\end{eqnarray}
where $X$, $Y$, and $Z$ are laboratory-fixed axes.
Similarly to Table I, the double and single indices correspond to the two incident photons and to the 
scattered photon, respectively. 
The value of $a^\zeta$ depends on the mode $\zeta$ and it is different for TO and LO. 
In the extreme case, where $\bar\beta^\zeta$ fully dominates the scattering, one should observe a constant depolarization ratio
$\rho_{coh}=(\bar\beta_{HVV}^\zeta/\bar\beta_{VVV}^\zeta)^2=1/9$.

\begin{figure}[h]
\includegraphics[width=7cm]{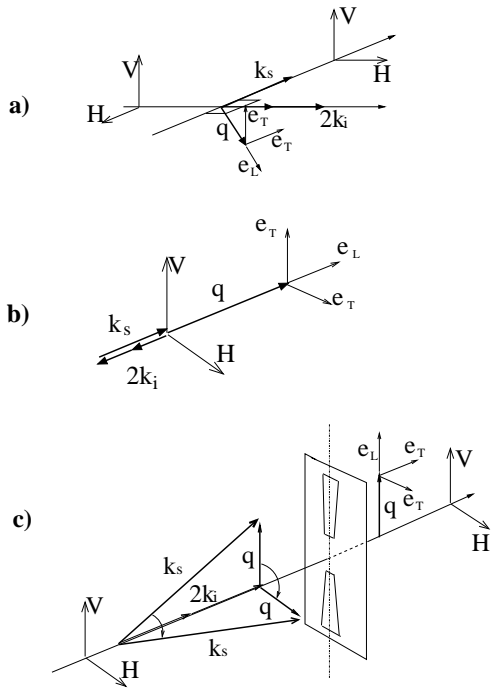}
\caption{HRS scattering geometries and polarization of the TO (${\bf e}_T$) 
and LO (${\bf e}_L$) components:
{\bf a)} 90$^\circ$-scattering, {\bf b)} backscattering,  {\bf c)} near-forward
scattering. For the latter, the vertical diaphragm is used to select only the LOs in a VV spectrum. } 
\end{figure}
Finally, the scattered intensity of these collective waves also depends on the
wave vector {\bf q}, as shown on Fig. 4.
The selection rules depend on the scattering geometry. 
From these pictures, and the tensor elements (7), one easily calculates the expected intensities 
for the TO and LO modes in the perfect $\infty\infty m$ symmetry.
These are summarized in Table II, where one sees that both TO and LO 
scatter at 90$^\circ$, while only TO modes are active in backscattering. 
\begin{table}[t]
\caption{Selection rules for TO and LO modes in the average ($\infty\infty m$) symmetry goup.}
\begin{tabular}{|c|c|c|c|c|}\hline
Scat. geometry & I$_{VV}$ & I$_{VH}$ & I$_{HV}$ & I$_{HH}$  \\
\hline\hline
90$^\circ$ & 9$a^2_{TO}$ & $\frac{1}{2}a^2_{TO}+\frac{1}{2}a^2_{LO}$ & $a^2_{TO}$ &
$\frac{1}{2}a^2_{TO}+\frac{1}{2}a^2_{LO}$ \\ \hline
180$^\circ$ & 9$a^2_{TO}$ & $a^2_{TO}$ & $a^2_{TO}$ & 9$a^2_{TO}$ \\ \hline
0$^\circ$ + vert. slit & 9$a^2_{LO}$ & $a^2_{TO}$ & $a^2_{LO}$ & 9$a^2_{TO}$ \\ \hline
\end{tabular}
\end{table}
In near-forward scattering, the wave vector {\bf q} is perpendicular to the optical axis.
It can take all orientations around this axis (Fig. 4c) and the scattering plane is undefined. 
To be able to define the polarization, it is necessary to restrict {\bf q} to a plane.
This is achieved by a slit placed at the exit pupil of the collecting lens (Fig. 4c).
A vertical slit defines a vertical scattering plane, while a horizontal slit gives a horizontal scattering plane.
In our implementation, the slit reduces the internal angular aperture to $\sim 4^\circ $ perpendicular to its direction, 
leading to an almost complete removal of the TO contribution from VV spectra in the configurations of Fig. 4c.
The reduction of the TO is estimated to be by a factor $\sim$ 25. 
This allows measuring the LO mode alone, with an intensity I$_{VV}\propto 9a_{LO}^2$.
Alternatively, the TO alone is observed in HH polarization with an intensity proportional to $9a_{TO}^2$.
This permits the measurement of the LO/TO intensity ratio.

Returning to B$_2$O$_3$, there are two types of IR active modes in the D$_{3h}$ point group:
the polar vibrations of symmetry A$''_2$ and those of symmetry E'.
The A$''_2$ modes correspond to out-of-plane motions. 
One eigenvector is a translation of the rigid ESUs which are charged (T$_z$), while the other is an internal vibration with 
displacements of the B and O atoms in opposite directions (see Table I).
The E' modes are in-plane motions. 
One representation corresponds to translations of the rigid ESUs (T$_x$, T$_y$), 
while the others are internal vibrations, as shown for example in Table I. 
These two types of modes are associated with the broad bands centered at $\sim$ 700 cm$^{-1}$ and $\sim$ 1400 cm$^{-1}$, 
respectively.
These modes are naturally present in IR absorption (TO) and IR reflectivity (LO), as shown in Fig. 1.   
Small frequency differences in the positions of the main maxima observed in different spectroscopies
could partly originate from differences in the samples. 
However, one cannot exclude an intrinsic property arising from the  
specificities of IR and HRS spectroscopies, which are sensitive to
fluctuations of first and third rank tensors, respectively  \cite{Den87}.
In order to determine the symmetry of the modes on the basis of the analysis above, we performed 90$^\circ$-, 180$^\circ$-,
and 0$^\circ$-scattering spectroscopies of the A$''_2$ and E' polar bands of $v$-B$_2$O$_3$. 

\subsection{Out-of-plane displacements A$''_2$}

The restoring forces for out-of-plane displacements are significantly smaller than for in-plane ones.
According to Ref. \cite{Han93}, the HRS band at $\sim 700$ cm$^{-1}$ can thus be attributed to A$''_2$ modes. 
\begin{figure}[h]
\includegraphics[width=8cm]{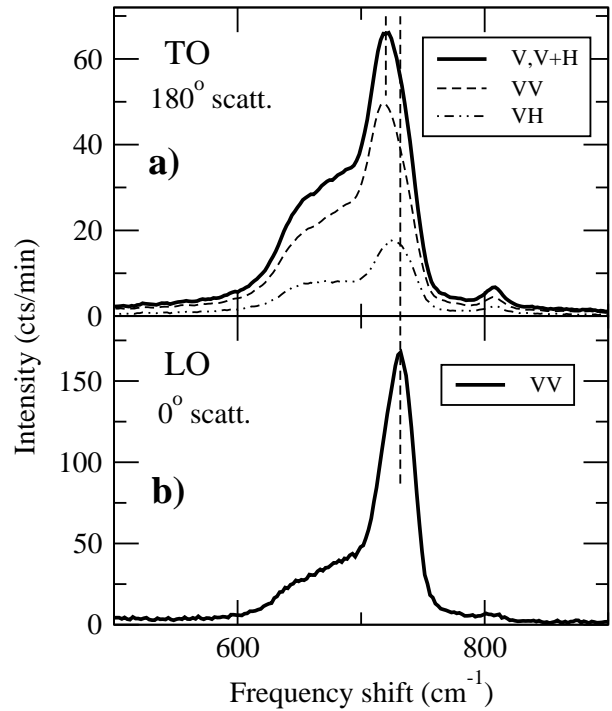}
\caption{Decomposition of the {\bf a)} TO and {\bf b)} LO responses for 
the A$''_2$ vibrations. The LO components (V,V) has been obtained in forward
scattering with a {\it vertical} slit, as explained in the text.}
\end{figure}
Fig. 5 shows separately the TO and LO contributions for these modes. 
As only the TOs scatter in backscattering, it is straightforward to extract their depolarization ratio. 
We find  $\rho^{A''_2}_{exp} = 0.36\pm 0.01$ at 720 cm$^{-1}$, far from the value 1/9 expected for scattering
arising only from $\bar{\bm\beta}$.
This type of discrepancy was observed in all the glasses that we investigated so far.\cite{Sim07} 
We believe it originates from the local term $\Delta\beta^\zeta$ which also contributes to scattering. 
The depolarization ratio slightly depends on the frequency, as revealed by the different profiles of the VV and VH spectra in Fig. 5a. 
The TO clearly exhibits a double-peak structure with a broad component centered around 665 cm$^{-1}$ and a more intense narrow one at 720 cm$^{-1}$. 
It is known from IR spectroscopy (see Fig. 1) that the LO mode is not far from the TO one. 
In that case, the LO can be measured alone by
performing HRS in near forward scattering (Fig. 4c). 
The doublet is also seen in this LO response presented in Fig. 5b.  
The maximum of the sharp LO component is at 731 cm$^{-1}$, 
which gives a small TO-LO splitting of only 11 cm$^{-1}$ for this particular spectral component.  
A significant value for the TO-LO splitting cannot be extracted for the 
broad component. 

To gain further information it is now interesting to compare $v$-B$_2$O$_3$  
to the lithium borate glasses 4B$_2$O$_3$:Li$_2$O and  2B$_2$O$_3$:Li$_2$O.
A major effect of lithium in boron oxide is to produce BO$_4$ tetrahedra for charge compensation.
This dramatically reduces the concentration of boroxol rings.
The disappearance of the A$'_1$ breathing mode at 808 cm$^{-1}$ in Fig. 6b is a clear signature for this.
As the number of rings is reduced with increasing Li content, 
the two maxima in Fig. 6b transform into a single peak at an intermediate frequency of $\approx 696$ cm$^{-1}$. 
Hence, we associate the sharp line of B$_2$O$_3$ at $\approx 720$ cm$^{-1}$ to 
A$''_2$ vibrations of rings, and the broader contribution at 665 cm$^{-1}$ to A$''_2$ vibrations of triangles.
This trend has been recently confirmed by first principle calculations:
making the rings silent in the simulation shifts the position of the A$''_2$ peak
to lower frequencies \cite{PasPrivate}.
\begin{figure}[h]
\includegraphics[width=8.5cm]{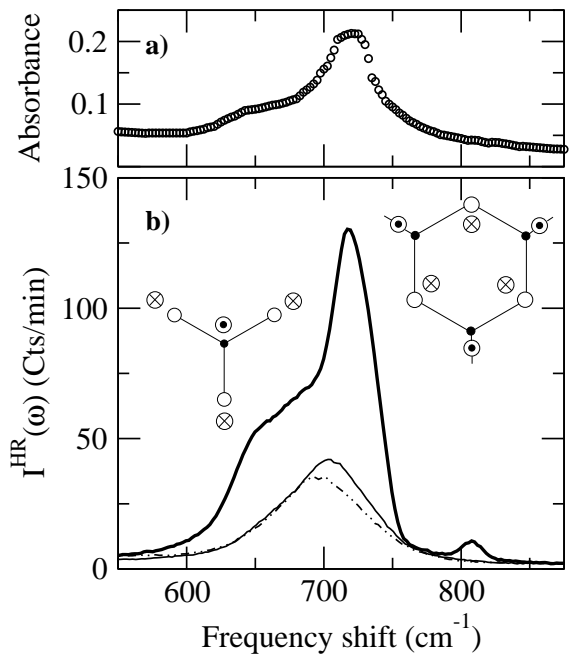}
\caption{{\bf a)}  A$''_2$ polar band in B$_2$O$_3$ as obtained from IR transmission \cite{Ten72}. 
{\bf b)} HRS VV spectra from the A$''_2$ polar band in borate glasses obtained in 90$^\circ$-scattering: 
``{\bf ---}'' $v$-B$_2$O$_3$, ``---'' 4B$_2$O$_3$:Li$_2$O, and 
``$\cdot\cdot\;$--$\;\cdot\cdot$'' 2B$_2$O$_3$:Li$_2$O. }  
\end{figure}
The eigenvectors of the A$''_2$ vibration for triangles and rings are shown in Fig. 6b. 
Returning to Fig. 1, one notes that the A$''_2$ vibrations are 
also observed in RS, although they should be forbidden by symmetry for perfect ESUs (see Table I). 
However, this neglects center-of-mass motion which is Raman active.
Also, a small departure from perfect planarity induces a permanent dipole moment 
that very effectively relaxes the Raman selection rule.
Thus, the RS signal could arise from vibrations of non-planar ESUs. 
The rings, being more regular than the triangles, should then give a weaker relative intensity in RS than in IR and HRS, which agrees
with observations in Fig. 1c. 

\subsection{In-plane displacements E$'$}

The weaker HRS efficiency, and the many components belonging to the band centered at 
$\sim$ 1400 cm$^{-1}$, do not allow for such detailed quantitative analysis as done for the A$''_2$ band.
In that case the LO is clearly separated from the TO (Fig. 1a) and not much would be gained by showing our measurement similar
to that in Fig. 5b. 
\begin{figure}[h]
\includegraphics[width=8.5cm]{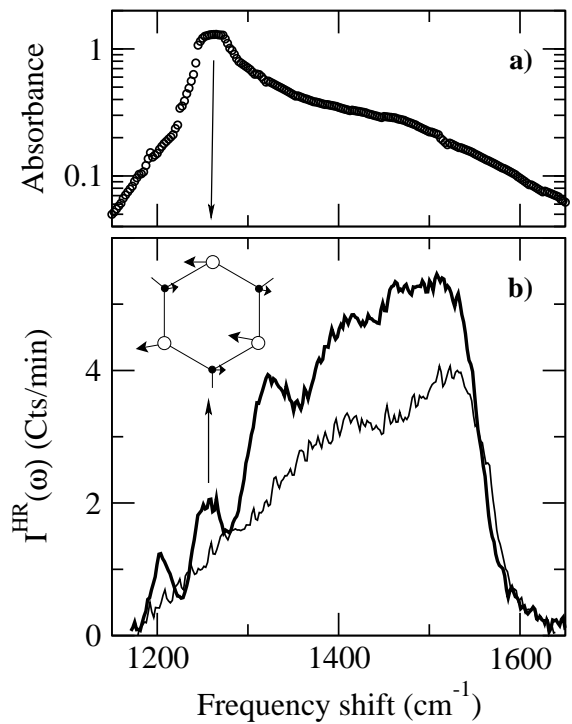}
\caption{{\bf a)}  E' polar modes of $v$-B$_2$O$_3$ as obtained from IR transmission \cite{Ten72} on a semi-logarithmic scale. 
{\bf b)} 90$^\circ$-scattering hyper-Raman spectra (VV) from the E' polar modes of  $v$-B$_2$O$_3$ (bold line) and 
4B$_2$O$_3$:Li$_2$O (thin line).}  
\end{figure}
Fig. 7b zooms on the 1400 cm$^{-1}$ frequency region in a geometry which does not scatter the LO. 
Unlike A$''_2$ modes, these high frequency vibrations exhibit a completely different shape in HRS and IR.  
The main IR line at 1265 cm$^{-1}$ is found in HRS, but with a rather weak intensity.
Its depolarization ratio is roughly estimated to be $\rho_{HR}= 0.37\pm 0.05$, again far from 1/9 but also 
smaller than the value 2/3 that would apply to non-polar displacements.
This confirms that this mode is polar in nature.
It is also expected to be active in RS (Table I), with a depolarization ratio of 3/4, which is observed in experiment. 
The activity of the line at 1265 cm$^{-1}$ in IR, RS, and HRS suggests an internal polar vibration E$'$ of the D$_{3h}$ 
point group (Table 1).
The mode disapears in the lithium borate glass 4B$_2$O$_3$:Li$_2$O (Fig. 6b), confirming \cite{Has92} that it is associated 
with motions in rings.
As it dominates the IR absorption spectra, it must correspond to displacements with the largest dipole moment 
fluctuations. 
From isotopic substitution, it was shown that this mode mainly involves the oxygen atoms \cite{Win82}.
This presumably also implies coupling with the center of mass motions (T$_x$, T$_y$).  
A tentative eigenvector is sketched in Fig. 7b. 

In addition, the HRS spectra exhibit 
two narrow lines at 1210 cm$^{-1}$ and 1325 cm$^{-1}$, plus two broad bands at higher frequencies, $\sim$ 1400 and $\sim$1500 cm$^{-1}$. 
The three peaks on the low frequency side clearly disappear in the lithium borate HRS spectrum of Fig. 7b.
We conclude that the narrow lines are associated with motions in boroxols, and that the broad bands correspond to motions in triangles. 
This confirms previous RS assignments \cite{Has92}. 
Finally, the comparison between these two spectra possibly reveals another ring vibration around 1460 cm$^{-1}$.
In any case, there seems to be sufficient richness in this band to fully account for the five E$'$ internal vibrations listed in (1).

\subsection{Polaritons in near-forward scattering}

Polaritons originate from the electro-mechanical coupling between transverse polar modes and the incident electromagnetic field.
This photon-phonon interaction requires that both the frequencies and wave vectors of both types of modes match, which
occurs at small scattering angles $\sim \omega _{\rm TO}/\omega _{\rm i}$, {\em i.e.} in near-forward scattering.
Owing to this interaction, the phase velocity of transverse polarization waves cannot exceed the speed of light in the medium.
In consequence, for scattering angles decreasing typically below 10$^\circ $, the frequency of the TO modes 
decreases down to the nearest LO component, with a slope smaller than that of the photon branch \cite{Poulet}.
The observation of polaritons by HRS was pionnered by Denisov {\it et al} \cite{Den87}.
A detailed analysis of their properties in borate glasses is not the purpose of this paper. 
However, the observation of this coupling can be very helpful to determine the polar character of vibrations.
\begin{figure}[h]
\includegraphics[width=8.5cm]{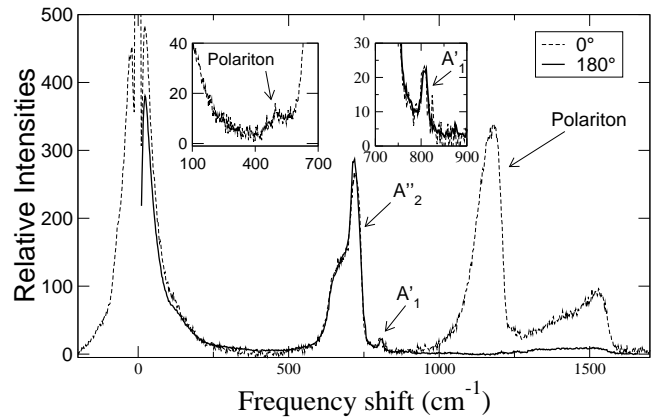}
\caption{Comparison between near-forward and backscattering HRS spectra with VV analysis. 
The former has been obtained using a {\it horizontal} slit in order to select only the TO polar components. 
The left inset zooms on the small polariton associated with the A$''_2$ TO band. 
The two spectra have been normalized in intensity to the maximum of the incoherent A$'_1$ mode shown in the right inset.}  
\end{figure}
The VV spectrum presented in Fig. 8 (thin line) has been obtained in near 
forward scattering  with  angles $\theta$ in the sample integrated from 
4$^\circ $ to 20$^\circ $, and a horizontal slit in order to reject the LO modes. 
A strong polariton is clearly visible around 1180 cm$^{-1}$. 
It is presumably associated with the main polar E' vibration at 1265 cm$^{-1}$. 
In contrast, the polariton associated with the polar A$''_2$ modes is very weak.
It is shown by the arrow in the inset. 
For this scattering geometry the TOs are active with an intensity $I\propto 9a^2_{TO}$. 
The result is compared with the backscattering HRS response obtained in VV (bold line) which also measures only the TO modes with $9a^2_{TO}$. 
As an incoherent and non-polar scatterer, the intensity of the A$'_1$ mode does not depend on the wave vector ${\bf q}$. 
Owing to this, it is possible to scale the spectra at 0$^\circ $ and 180$^\circ $ to the maximum of this mode.  
The observed differences should essentially arise from the polariton activity. 
This unambiguously shows that there exists polaritons in the entire broad band from $\sim$ 1200 to $\sim$ 1600 cm$^{-1}$.
A detailed analysis of this complex band could presumably only be achieved with careful simulations.  

Looking at the boson peak region, one observes an intensity excess in near-forward scattering.
This would also deserve a more detailed analysis. 
This ${\bf q}$-dependence of the intensity suggests a small polar component in the BP.
However, our data at 90$^\circ $ and 180$^\circ $ are very similar and the polarization 
analysis gives the same depolarization ratio $\rho=0.60$ at the BP maximum for both scattering 
geometries, \cite{Sim07} while the LOs are active in 90$^\circ $ (VH) and silent in backscattering. 
We conclude from the above that there exist motions in the BP of B$_2$O$_3$
which induce fluctuations of the dipolar moment, but that the dominant modes in HRS from the BP are
non-polar in nature.

\section {The boson peak}

The last main feature in HRS is the boson peak (BP) at low frequencies.
Following our model, it is unlikely that the microscopic origin of the BP relates to any {\em internal} vibration of
the D$_{3h}$ point group as they have mostly been assigned in former Sections.
However, there remains {\it external} vibrations, the translations ({\bf T}) and the rotations ({\bf R}) of rigid ESUs.
These motions are non-genuine vibrations for isolated molecules.\cite{Her45} 
They are characterized by diffusive relaxation in the liquid phase and transform into real vibrations,
usually at low frequencies, in the glass phase. 
Hence, one should use the term ``libration'' to designate angular oscillations of the ESUs in the latter case, as the
mean equilibrium orientations are then fixed.
For what translations are concerned, they should not be confused with long wave acoustic modes.
The latter are collective waves whose HRS is strictly forbidden in the $\infty \infty m$ symmetry of the glass phase.\cite{Den87}
However quasi-local relative displacements of nearby ESUs can be active in HRS from glasses just as optic modes 
are in crystals.
\begin{figure}[h]
\includegraphics[width=8cm]{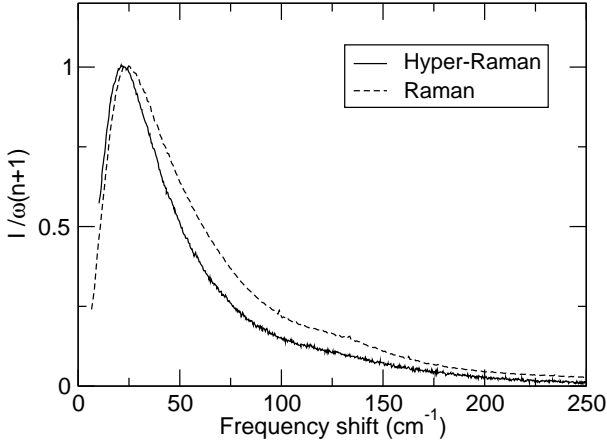}
\caption{VH spectra of the RS (- -) and HRS (---) boson peaks. 
The data are normalized to the BP maximum.  }
\end{figure}

Figure 9 shows that HRS and RS BPs, although not identical, are very similar. 
It suggests that the BP modes are active in both spectroscopies. 
It has been shown, using lithium borate glasses, {\it e.g.} in Ref. \cite{Bar00},  
that the RS BP of $v$-B$_2$O$_3$ is dominated by boroxol-ring motions rather than by triangle motions. 
The same applies to HRS as we observed a strong decrease in the intensity of the BP in lithium borate samples. 
This further supports the idea that the same modes contribute to the RS and HRS BPs.     
Proceeding by elimination, and taking into account that in-plane 
librations (A$'_2$) are hard motions silent in RS, the only remaining possibility for 
the BP is out-of-plane librations of rings (E" modes) coupled to their relative translations.
The translations are active in IR (Table I) and this could
explain the excess signal on the BP that is observed in near-forward scattering (Fig. 8). 
However, it has been shown in liquids that the low frequency HRS signal is mainly dominated by the rotational component of the external motions 
rather than by translational ones \cite{She02,Kaa96}. 
For example, liquid CCl$_4$ is composed of tetrahedra belonging to the point group T$_d$.
The strong relaxational hyper-Rayleigh line of CCl$_4$ was analyzed as the sum of a narrow and a broad lorentzian.
The narrow component was interpreted in terms of rotational diffusion and the broad one was attributed to {\it collective} 
intermolecular orientational motions \cite{Kaa96}. 
It is intuitive that in network glasses, rigid translations or librations of one ESU do couple with neighbouring ESU motions. 
Depending on the nature of the relative displacements of the structural units, 
this can lead to constructive or destructive coherent effects in scattering.
\begin{figure}[h]
\includegraphics[width=8cm]{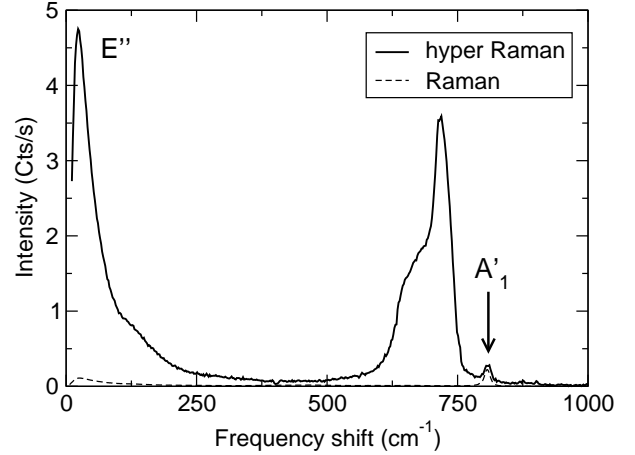}
\caption{Raman and hyper-Raman spectra of $v$-B$_2$O$_3$. The intensities
are normalized to the area of the A'$_1$ mode active in both spectroscopies.
This presentation emphasizes the comparatively strong efficiency of the BP in
HRS.}  
\end{figure}
If constructive coherence occurs, it can lead to an anomalously strong HRS efficiency, which is indeed observed as seen in Fig. 10.
The intensity of the BP is comparable that of the strongest polar vibrations which are coherent scatterers in HRS.
Incoherent non-polar vibrations give a relatively weak HRS signal, as shown by the A$'_1$ mode discussed in Section IV.
The scattering efficiency of the BP relative to the breathing A$'_1$-mode is much larger in HRS than in RS.
This is emphasized by the normalization used for Fig. 10. 
\begin{figure}[h]
\includegraphics[width=8.5cm]{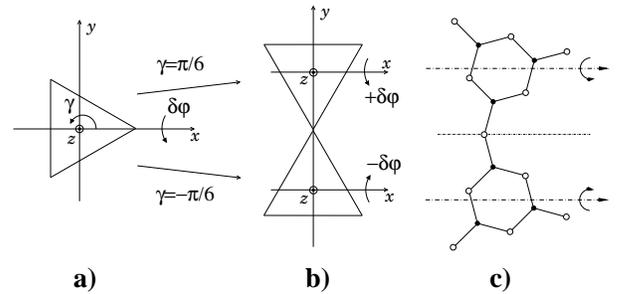}
\caption{Definition of the axes and rotation angles for a single ESU (a) and for a particular
out-of-plane libration of two adjacent ESUs (b); (c) shows a possible ESU configuration approximating (b).}  
\end{figure}

Coherent effects on the RS and HRS BP relate to the symmetries of the
polarizability and hyper-polarizability tensors of the E" modes.
To show how this can occur, we present now a simple calculation for a particular coherent libration of a pair of ESUs.
In the D$_{3h}$ point group, both BO$_3$ and B$_3$O$_3$ rings transform like equilateral triangles (Fig. 11).
For the boroxols, the B atoms point to the apices of these triangles owing to their bonding to external O atoms.
The Raman susceptibility ${\bm \chi}^{(1)}$ can be decomposed into its isotropic and anisotropic parts.
We neglect the modification of the isotropic tensor by librations, as this must be a second order effect.
The traceless anisotropic part has elements
$\chi^{(1)}_{11}=\chi^{(1)}_{22}= - \frac{1}{2} \chi^{(1)}_{33}$. For convenience we write $\chi^{(1)}_{11}\equiv - c/3$.
Now consider that the ESU of Fig. 11a is rotated by a finite angle $\gamma $ around the $z$-axis and that it librates by an angle 
$\delta \varphi$ around the initial $x$-axis.
The in-plane rotation by $\gamma $ does not modify ${\bm \chi}^{(1)}$, while the out-of-plane libration around $x$ modulates it by
\begin{equation}
\delta {\bm \chi}_{x}^{(1)}= \delta\varphi \left |  \begin{tabular}{ccc}
$\cdot$ & $\;\;\cdot$ & $\cdot$ \\
$\cdot$ & $\;\;\cdot$ & $\;\;c$ \\
$\cdot$ & $\;\;c$ & $\cdot$ \\
\end{tabular}\right | \;\;\;\;\; ,
\end{equation}
One should note that, similarly, a libration by $\delta \phi$ around the $y$-axis gives
\begin{equation}
\delta {\bm \chi}_{y}^{(1)}= \delta\phi \left |  \begin{tabular}{ccc}
$\cdot$ & $\;\cdot$ & $\;-c$ \\
$\cdot$ & $\;\cdot$ & $\cdot$ \\
$-c$ & $\;\cdot$ & $\cdot$ \\
\end{tabular}\right | \;\;\;\;\; .
\end{equation}
One recognizes then in Eqs. (8) and (9) the two tensors transforming like R$_x$ and R$_y$ in the last entry of Table I.

Consider now the hyper-Raman second-order susceptibility tensor ${\bm \chi}^{(2)}$.
With the definition of axes in Fig. 11a, the only nonvanishing elements are   
$\chi^{(2)}_{111} = b = \chi^{(2)}_{11} $, 
$\chi^{(2)}_{122} =  -b =  \chi^{(2)}_{12} $, and 
$\chi^{(2)}_{221} =  \chi^{(2)}_{212} = -b =  \chi^{(2)}_{26} $, where contracted indices\cite{Indices} are used in the last term
of the equalities.
The structure of that tensor is identical to that in the first entry of Table I.
We call it ${\bm \chi}^{(2)}_{\rm X}$, indicating that an apex of the triangle lies on the positive $x$ axis.
Now, if the triangle is rotated by $\gamma = \pi /2$, so that an apex lies on the positive y-axis, one finds
$\chi^{(2)}_{21} =\chi^{(2)}_{16} = b$ and $\chi^{(2)}_{22} = -b$. This is called ${\bm \chi}^{(2)}_{\rm Y}$, so that
\begin{equation}
{\bm \chi}_{\rm X}^{(2)}= \left \|  \begin{tabular}{cccccc}
$\; b$ & $-b$ & $\cdot$ & $\; \cdot$ & $\; \cdot$ & $\cdot$ \\ 
$\cdot$ &  $\cdot$ & $\cdot$ & $\; \cdot$ & $\; \cdot$ & $-b$ \\
$\cdot$ & $\cdot$ & $\cdot$ & $\; \cdot$ & $\; \cdot$ & $\cdot$ \\
\end{tabular} \right \| \;\; , \; \; 
{\bm \chi}_{\rm Y}^{(2)}= \left \|  \begin{tabular}{cccccc}
$\cdot$ & $\cdot$ & $\cdot$ & $\; \cdot$ & $\; \cdot$ & $\; b$ \\ 
$\; b$ &  $-b$ & $\cdot$ & $\; \cdot$ & $\; \cdot$ & $\cdot$ \\
$\cdot$ & $\cdot$ & $\cdot$ & $\; \cdot$ & $\; \cdot$ & $\cdot$ \\
\end{tabular} \right \|  \;  .
\end{equation}
Incidentally, the latter form corresponds to the choice of axes in Ref.\cite{Gio65}.
Now, if ${\bm \chi}^{(2)}$ is rotated by an arbitrary angle $\gamma $, one simply finds
\begin{equation}
{\bm \chi}_{\gamma}^{(2)} \; = \; \cos 3 \gamma \; {\bm \chi}_{\rm X}^{(2)} \; + \; \sin 3 \gamma \; {\bm \chi}_{\rm Y}^{(2)}  \; \; .
\end{equation}
Librating ${\bm \chi}_{\rm X}^{(2)}$ and ${\bm \chi}_{\rm Y}^{(2)}$ around the $x$-axis by $\delta \varphi $, one obtains
\begin{equation*}
\hspace{-3.5cm}\delta {\bm \chi}_{{\rm X},x}^{(2)}= \delta \varphi \left \|  \begin{tabular}{cccccc}
$\cdot$ & $\;\; \cdot$ & $\;\; \cdot$ & $\; b$ & $\; \cdot$ & $\; \cdot$ \\ 
$\cdot$ &  $\;\;\cdot$ & $\;\;\cdot$ & $\cdot$ & $b$ & $\cdot$ \\
$\cdot$ & $\;\;\cdot$ & $\;\;\cdot$ & $\cdot$ & $\cdot$ & $b$ \\
\end{tabular} \right \| \;\;\; , 
\end{equation*}
\begin{equation}
\qquad\qquad\qquad\delta {\bm \chi}_{{\rm Y},x}^{(2)}= -\delta \varphi \left \|  
\begin{tabular}{cccccc}
$\cdot$ & $ \cdot$ & $ \cdot$ & $\cdot$ & $-b$ & $\cdot$ \\ 
$\cdot$ &  $\cdot$ & $ \cdot$ & $\; b$ & $\cdot$ & $\cdot$ \\
$-b$ & $\; b$ & $\cdot$ & $\cdot$ & $\cdot$ & $\cdot$ \\
\end{tabular} \right \| \; .
\end{equation}
Again, one recognizes the tensors in the last row of Table I. For R$_x$ the identity is immediate.
For R$_y$ it results from
$\delta {\bm \chi}_{{\rm X},y}^{(2)}/\delta \phi = -\delta {\bm \chi}_{{\rm Y},x}^{(2)}/\delta \varphi $,
as R$_y$ corresponds to the libration of ${\bm \chi}_{\rm X}^{(2)}$ around $y$.
Finally, from (11), the fluctuating susceptibility for a libration around $x$ of an ESU rotated by $\gamma $ around $z$ simply is
\begin{equation}
\delta {\bm \chi}_{\gamma , x}^{(2)} \; = \; \cos 3 \gamma \; \delta {\bm \chi}_{{\rm X},x}^{(2)}  \; + 
\; \sin 3 \gamma \; \delta {\bm \chi}_{{\rm Y},x}^{(2)}   \; \; .
\end{equation}

With the above, one can now calculate the RS and HRS responses of {\em two} ESUs connected by a common apex that moves perpendicularly to the plane
of the drawing.
Fig. 11b show the case where the two ESUs are rotated from Fig. 11a by $\pm \pi / 6$.
The Raman response is independent of $\gamma $, and since one unit moves by $+\delta \varphi $ while the other moves by $-\delta \varphi $, the sum of
two expressions (8) gives a total $\delta {\bm \chi }^{(1)}$ which is strictly zero.
This is coherent cancellation.
For the HRS response, $\sin 3 \gamma = \pm 1$, so that $\sin 3 \gamma \; \delta \varphi  = \mid \delta \varphi \mid $ for both ESUs.
The HRS susceptibilities add, meaning that the total intensity is multiplied by 4.
This is perfect constructive coherence.

These opposite coherence effects arise from the general property that ${\bm \chi }^{(1)}$ is even under inversion, 
while ${\bm \chi }^{(2)}$ is odd.
For that reason the coherent cancellation of RS for two adjacent ESUs moving symmetrically out-of-phase will always occur.
Hence, one should generally expect that coherence negativelly affects the RS strength of rigid unit motions, while it is
likely to enhance the HRS strength.
In the present case, the observed RS BP might amount to just the incoherent contribution, or even less if the coherent cancellation
is very effective.
For what concerns the HRS strength, $\gamma = \pm \pi /6$ is the most favorable case in the simple model of Fig. 11b.
A realistic estimate of the HRS coherent enhancement should take into account that the libration axes of adjacent units need not be parallel.
Hence, it seems that only large simulations based on an accurate glass structure will eventually be able to 
reproduce the experimental result.
This would presumably turn out to be an exacting test for the accuracy of the model structure.
Very roughly speaking, if $N$ ESUs move in perfect coherence, one would expect a HRS enhancement $\propto N^2$.
From Fig. 9, we estimate the HRS BP to be 50 times more intense than the RS one.
{\em If} the latter is fully incoherent, and {\em if} the HRS and RS coupling strengths are comparable,
this could imply a value $N\sim 7$.
This very simple estimate makes it clear that this coherence question deserves further investigations.

As a final observation, one should note the slight difference in the maximum position of the RS and HRS BPs in Fig. 9.
We find  $\omega^{HR}_{BP}\simeq 23$ cm$^{-1}$, while  $\omega^{R}_{BP}\simeq 26$ cm$^{-1}$.  
The comparison with neutron data shows that the RS BP coincides very 
well with the {\it random phase} part\cite{Eng98} of the reduced vibrational density of states.\cite{Sim06}
This good agreement presumably results from the incoherent nature of the Raman scattering.
The lower position of the HRS BP could thus reflect the coherent enhancement, since HRS
would be more effective for coherent packets of larger size which on the average would librate at lower frequencies.

\section{Conclusion}

Hyper-Raman scattering has been investigated in $v$-B$_2$O$_3$.  
HRS spectra are very different from RS and IR ones, emphasizing the importance of selection rules. 
Our experimental results motivated an analysis in terms of elementary structural units (ESUs).
We assumed a simple structural model involving just one BO$_3$ triangle and one B$_3$O$_3$ boroxol ring. 
Both belong to the D$_{3h}$ point group.
The experiments show that the breathing vibration of boroxol rings (A'$_1$ mode at 808 cm$^{-1}$) is silent in IR, 
and active in RS and HRS, as expected from symmetry considerations.
Moreover, its HRS depolarization ratio matches the theoretical value of 2/3 expected for incoherent non-polar scatterers 
in the D$_{3h}$ point group.
The mode analysis is then applied to the other main features of the HRS spectra of $v$-B$_2$O$_3$.
This allowed identifying the two main polar bands of the IR spectra as 
the out-of-plane vibrations of triangles and rings around 720 cm$^{-1}$ (A"$_2$ modes) and as the in-plane vibrations 
of rings at 1265 cm$^{-1}$ (E' modes).
Proceeding by elimination, the BP is then attributed to out-of-plane {\it librational motions} of rigid boroxol
rings {\it coupled to their relative translations}. 
The latter are external vibrations of the ESUs that occur naturally at low frequency owing to the relatively weak 
restoring forces between ESUs.
Our conclusion for the boson peak parallels the concept of ``rigid unit modes'' (RUMs) proposed in Ref \cite{Tra98,Pal02}.
The spectroscopy reveals that the HRS activity of the BP mainly arises from the librational component (E'' symmetry) 
of these mixed modes. 
It has been established several times that the lowest optic-like modes are associated with librations  
of rigid units in glasses, {\it e.g.} $v$-SiO$_2$,\cite{Heh00,Buc86,Tar97,Gui97} Se,\cite{Ber94} and now in $v$-B$_2$O$_3$ (here
and in Ref. \cite{Sim06}). 
Internal E'' modes are found to be the lowest frequency  vibrations in  molecular ring systems in the gas state, 
such as boroxine H$_3$B$_3$O$_3$ \cite{Tos90}, or triazine
C$_3$N$_3$H$_3$ \cite{Mor97}.
Finally, the strong activity of the BP in HRS probably originates from the coherent motion of several adjacent ESUs.
A simple example shows how the scattering from two connected ESUs in an out-of-phase E'' motion adds in HRS while it subtracts in RS.
Considering the planar isotropy of the Raman tensor of D$_{3h}$
ESUs, symmetry suggests that on the average this effect should enhance the HRS efficiency compared to the RS one. 
Realistic calculations taking into account the structural disorder of the glass remain to be performed.
We do have preliminary HRS results on other glasses showing strong BPs and suggesting similar  
coherent enhancement.
As coherence is presumably related to medium range order, the information provided of HRS could become of considerable value.

The authors express their appreciation to Saint-Gobain Recherche, and in particular to Dr. M.-H. Chopinet and Dr. P. Lambremont, for
the availability of their glass preparation facility and for their guidance in obtaining suitable samples.
Thanks are also addressed to Dr. A. Matic from Chalmers University for providing samples of lithium borate glasses.


\begin{thebibliography}{99}
\bibitem{Indices} The notation $\chi^{(2)}_{ijk}$ is transformed to contracted indices $\chi^{(2)}_{im}$ with $m = j$ for $j = k$,
and $m = 9-(j+k)$ for $j\neq k$.
\end{thebibliography}
\end{document}